# Single Layer PCB Broadband Circular Polarisation Millimetre Wave Massive MIMO Array

Qian Xu[1], Yi Huang[2], Tian-Hong Loh[3], Manoj Stanley[2], Lei Xing[1], Min Wang[1], and Hui Gan[1]

[1] College of Electronic and Information Engineering, Nanjing University of Aeronautics and Astronautics, Nanjing, China, emxu@foxmail.com

[2] Department of Electrical Engineering and Electronics, The University of Liverpool, Liverpool, United Kingdom, yi.huang@liv.ac.uk

[3] Engineering, Materials & Electrical Science Department, National Physical Laboratory, Teddington, United Kingdom, tian.loh@npl.co.uk

*Abstract*—A single layer broadband millimetre wave (mm-wave) array with circular polarisation is proposed for the fifth generation (5G) mm-wave communication applications. The antenna element is designed to integrate the matching circuit into a single layer. Hence no additional matching and feeding networks are required which makes the antenna element very easy to synthesise a massive array. The array has a 10 dB return loss bandwidth between 26 GHz and 32 GHz. The polar angle axial ratio is less than 3 within the beamwidth of -60° to 60°. The proposed antenna structure is simple, low cost and robust, which we envisage suitable for the massive multiple-input and multiple-output (MIMO) applications.

*Index Terms*—circular polarisation, millimetre wave, antenna, MIMO.

## I. Introduction

Circularly polarised (CP) antennas are widely used in line-of-sight (LoS) communications, as it is not sensitive to the polarisation of the receiver and the LoS signal can be distinguished from other reflected signals with opposite polarisation [1]. A multilayer antenna has been proposed in [2] with two ports which have good isolations and the antenna can be configured to operate either with vertical or horizontal polarisation. Multilayer can indeed provide more freedom for the antenna design, and a wideband CP antenna with multilayer has been proposed in [3]. On the other hand, a single layer design is simple, low cost, easy to fabricate and more reliable in complex environment [4], especially for conformal structures. A coplanar waveguide feed CP slot antenna has been proposed in [5], and the main radiation is in both sides of the antenna. In this paper, we propose a novel design that combining all the above features, namely single layer, broadband, circular polarisation as well as single side radiation pattern.

The design philosophy and the proposed structure are explained in Section II, the simulated results are given in Section III, and discussion and conclusions are detailed in Section IV.

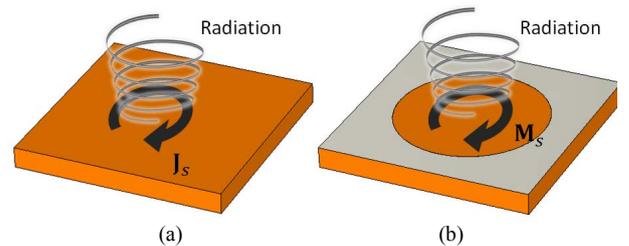

Fig. 1  The source of circular polarisation waves, (a) electrical current, (b) magnetic current.

## II. Proposed Design

### A. Design Philosophy

It is well known that there are mainly two types of CP antennas [6]: the resonator antenna and the travelling-wave antenna. A typical resonant antenna is the patch antenna, to realise a circular polarisation, the patch normally supports two orthogonal fields/modes of equal amplitude but in-phase quadrature (i.e. 90 degrees difference) [6]. If one port is not enough to excite the orthogonal modes, then two ports with a power divider and a phase shifter are required. For the travelling-wave antenna, the waves propagate along the transmission line and radiate simultaneously, providing a broad bandwidth (e.g. helical antenna [7]).

To realise a single layer, broadband and CP antenna, the travelling-wave type is preferred. As the resonant type antennas are relatively narrow band and sensitive to the geometric parameters and material properties, which means the manufacturing process needs to be controlled very carefully. The advantage of the resonant type antenna is that the electrical size can be very small (because of the Chu-Harrington limit [8]).

As can be seen in Fig. 1, an intuitive method to generate CP radiation is to have electrical current $\mathbf{J}_s$ and/or magnetic current $\mathbf{M}_s$ flow circularly and radiate simultaneously. Figure 1(b) is preferred as it has a cavity for each antenna element and could provide better isolations. Thus the problem to address is how to design a structure in the cavity in Fig. 1(b) to guide the current flows circularly and match the impedance within the desired working frequency range.

RO4350B, the relative permittivity is 3.48 and the thickness is chosen as 1.52 mm. To show the inner structure clearly, the substrate is hidden in Fig. 2(b) and an enlarged bird-eye view picture is illustrated in Fig. 2(c). The dimensions of the bottom layer and the top layer structures are given in Fig. 2(e) - (f). The antenna can be considered as a curved planar inverted-F antenna with a parasitic element and a matching circuit. By using the well-developed printed circuit board (PCB) technology, a cavity with shunted vias can be created, and the antenna is fed by a 50 Ohm K-type connector with a diameter of 1.27 mm of the inner conductor and 2.92 mm of the outer conductor.

By driving the current flow circularly and matching the impedance, a broadband CP antenna is realised. The antenna and the impedance matching circuit are integrated which simplifies the feeding network for an antenna array.

### III. Antenna Performance

The simulated results are detailed in this section which include *S*-parameters, current distributions, radiation patterns and axial ratios.

#### A. S-parameters

The simulated *S*-parameter of a single antenna element is given in Fig. 3(a), and a 4 × 4 elements array in Fig. 3(b) is also simulated and the mutual coupling *S*-parameters are shown in Fig. 3(c). As can be seen, the $S_{11}$ is smaller than –10 dB in the working frequency of 26 GHz – 32 GHz, but the mutual coupling between the ports is higher than the values in [2] (typically < –30 dB). However, the distance between the antenna elements is very compact, which is 4.9 mm ($\lambda/2$ at 30 GHz). To reduce the mutual coupling, the distance between the elements can be increased, the mutual coupling can be reduced to < –20 dB when the distance between elements is 9.8 mm, as shown in Fig. 3(d). The tradeoff is the scanning range of the antenna array. Another method is to use a few elements to form a subarray which is also a very practical solution.

#### B. Field Distributions

The mechanism of the circular polarisation can be verified from the current distribution and the electrical field distribution. The surface current density is shown in Fig. 4(a) – (d) with different phases; the corresponding E-field distributions are illustrated in Fig. 4(e) – (g), where the circularly polarised wave can be observed intuitively. As expected, the current is driven to flow circularly and radiate simultaneously.

#### C. Radiation Patterns

The patterns of the realised gain (including the mismatch of the antenna) are given in Fig. 5(a) – (f), the maximum gain is about 5.3 dB and the 3 dB beamwidth is about 90°. The radiation patterns in the working frequency are very symmetric about *z*-axis. The synthesised pattern with all elements excited simultaneously has also been

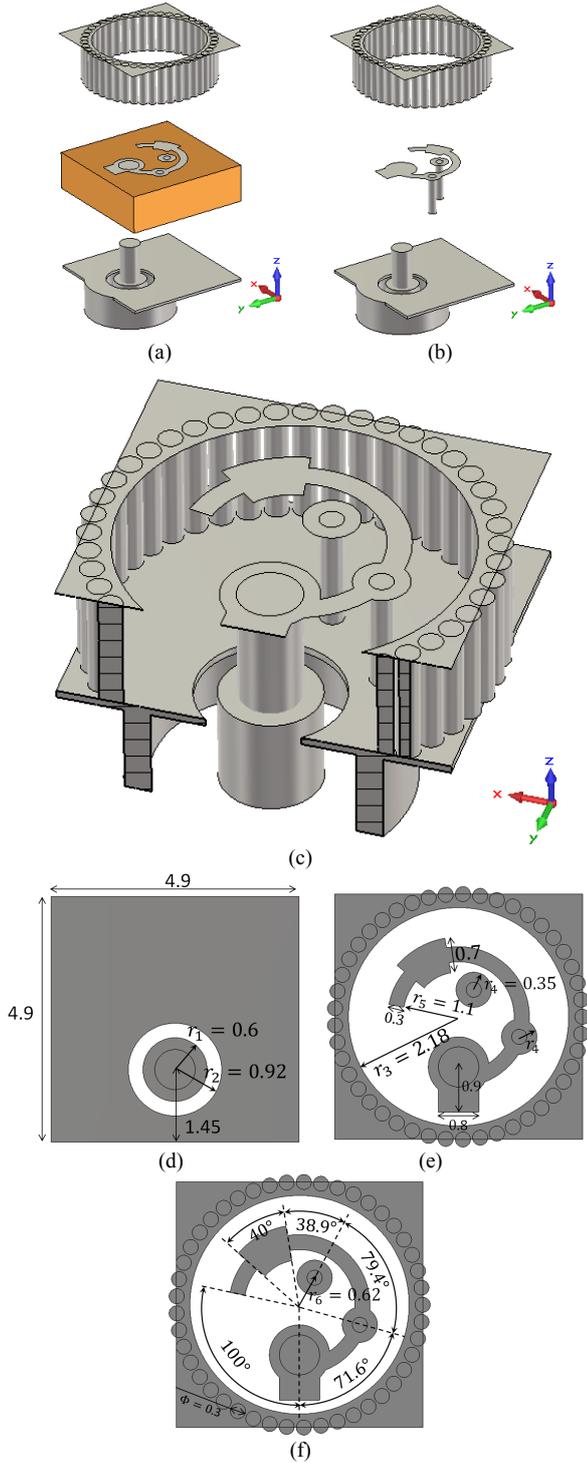

Fig. 2 The proposed radiation element, (a) exploded view of the antenna with dielectric substrate, (b) exploded view of the antenna with substrate hidden, (c) bird-eye 3D structure view of the antenna with substrate hidden, (d) bottom layer view, (e) and (f) top layer view, two figures are used to show the dimensions due to the limited space. Units: mm.

#### B. Proposed Structure

The proposed antenna element of this paper is shown in Fig. 2. The parameters of the antenna structure have been tuned and optimized. The substrate we use is Rogers

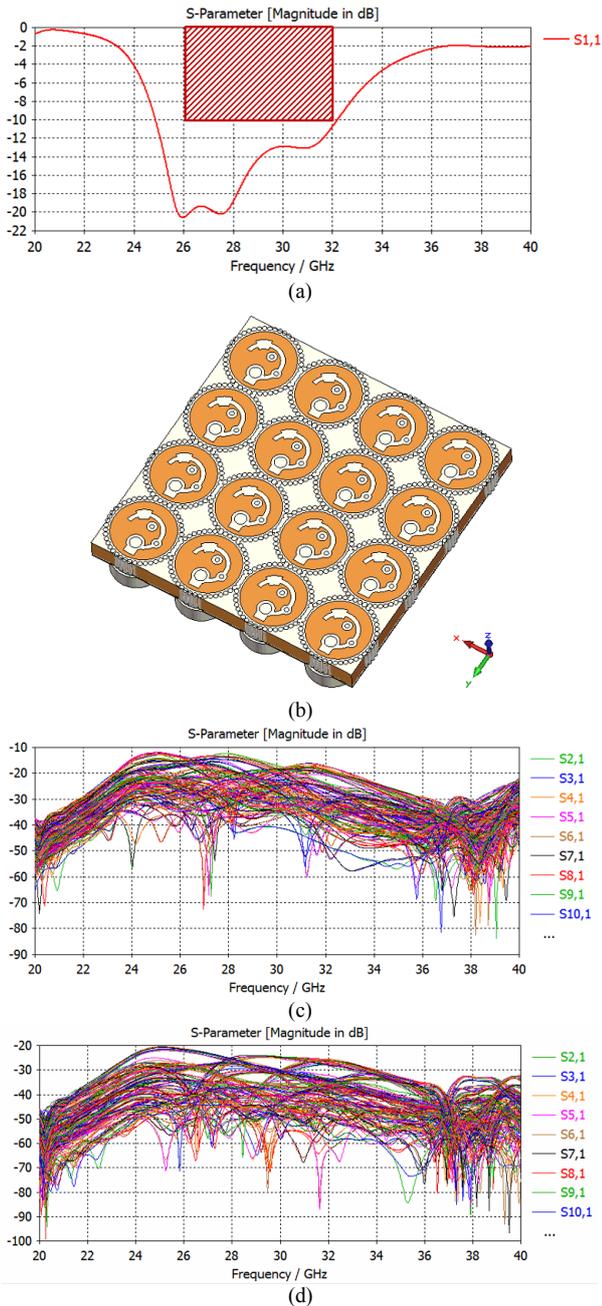

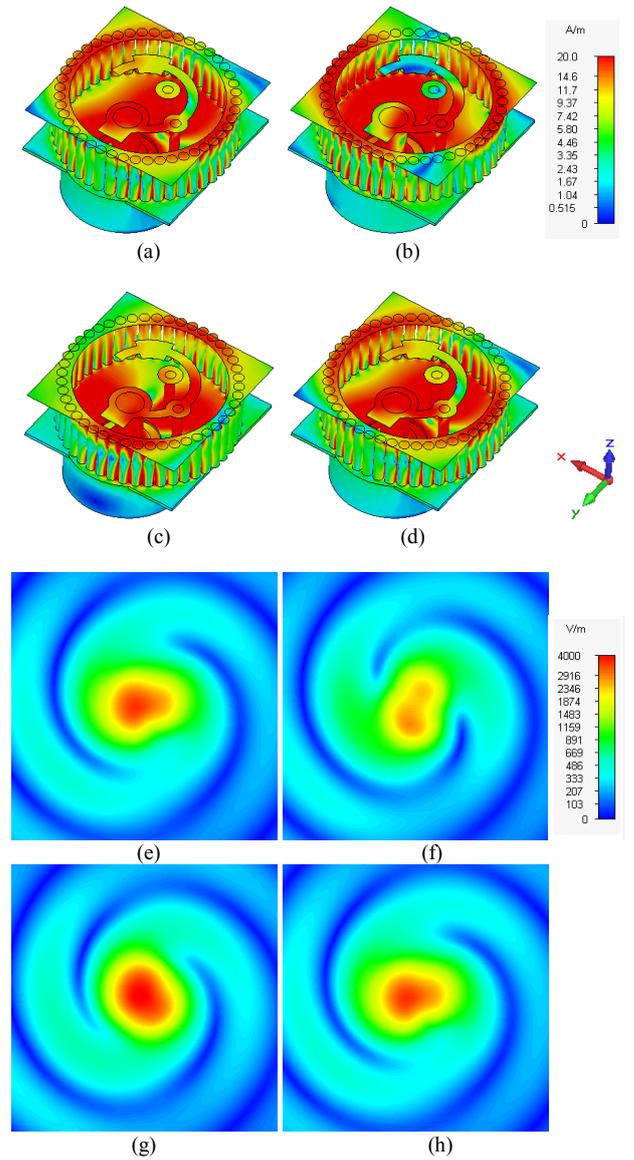

Fig. 3 (a) $S_{11}$ of the antenna element; (b) a 4 × 4 array model with a period of 4.9 mm; (c) mutual coupling between antenna elements of the simulated model in (b); (d) mutual coupling between antenna elements when the period of the array is 9.8 mm. Note that there is a total of 240 transmission loss S-parameter results presented each plot shown in Fig. 3(c) and (d).

Fig. 4 Current distribution and E-field distribution at 30 GHz, the E-field distribution is taken in a *xoy* cut plane with the height of 3 mm from the top of the antenna: (a) current distribution with the phase of 0°, (b) 60°, (c) 120°, (d) 180°; (e) the magnitude distribution of E-field tangential component with the phase of 0°, (f) 60°, (g) 120°, (h) 180°.

simulated and given in Fig. 5(g), which has a realised gain larger than 15.5 dB in the working frquency band.

### D. Axial Ratios

The dependency of axial ratios (ARs) with frequencies and angles are shown in Fig. 6(a) – (c). As can be seen, the AR is wide band in frequencies and wide beam in angles, the ARs are less than 3 in the polar angle of –60° ~ +60° in the frequency range of 27 GHz ~ 31 GHz. The 4 × 4 array has also been verified, the AR is given in Fig. 6(d). It is interesting to note that there is a slightly difference between Fig. 6(a) and Fig. 6(d), this is due to the difference between the embedded pattern and the pattern in free space of a single element. For the antenna in this paper, this effect is very small as each antenna is in a cavity and isolated with vias.

## IV. DISCUSSION AND CONCLUSIONS

A broadband circular polarisation mm-wave antenna element has been proposed, which is suitable for the massive MIMO applications. The antenna and the matching circuit are integrated in a cavity which can be formed by

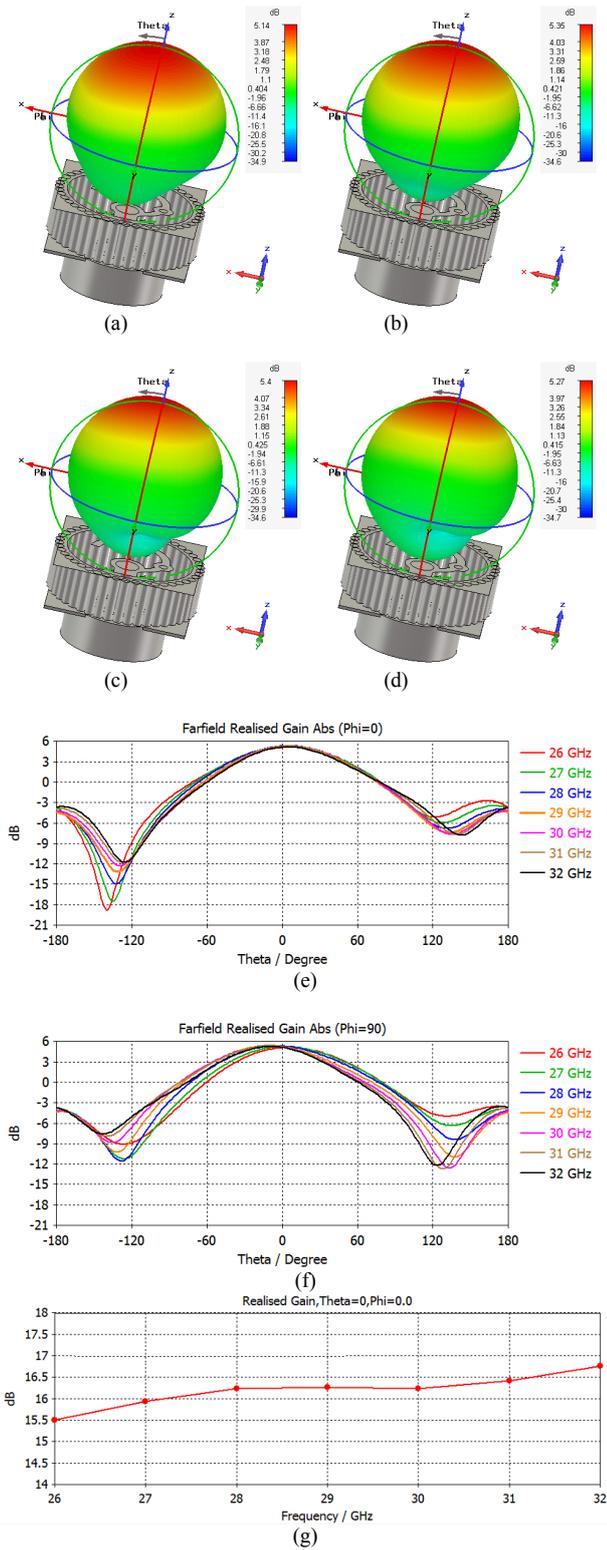
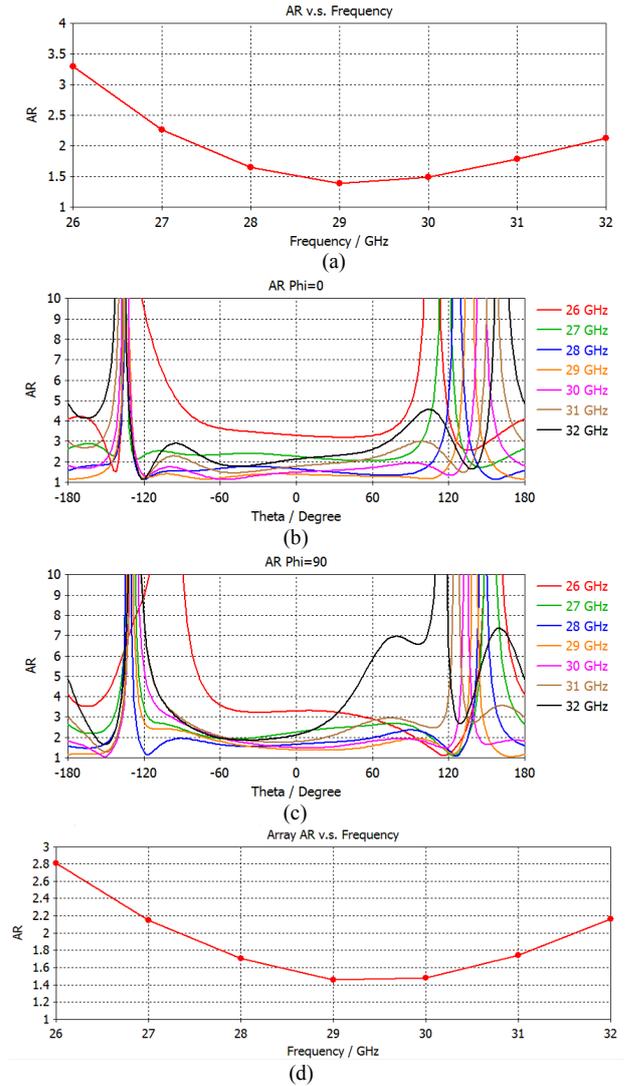

Fig. 5 The realised gain of the antenna element, (a) 26 GHz, (b) 28 GHz, (c) 30 GHz, (d) 32 GHz, (e) cut plane with azimuth angle phi=0°, (f) cut plane with phi=90°, (g) the realised gain of a 4 × 4 array with all elements excited in phase.

using vias and the ground plane. The working frequency is 26 GHz to 32 GHz, with a good AR bandwidth in frequency and beamwidth in angle. The performance of a

Fig. 6 (a) AR at direction $\theta = 0°, \varphi = 0°$ with different frequencies, (b) AR in the cut plane of $\varphi = 0°$, (c) AR in the cut plane of $\varphi = 90°$, (d) AR at $\theta = 0°, \varphi = 0°$ of the 4 × 4 array when all elements are excited in phase; the units are in linear scale.

4 × 4 array has also been simulated, which shows good results as well.

The mutual coupling of the antenna elements is not very low when the period of the array is about $\lambda/2$. However, the effect of mutual coupling could be reduced by using spatial diversity and subarray technology. Because the antenna structure is simple and can be made by using only one layer PCB technology, the low cost advantage could offer a good solution for antenna designs in massive MIMO applications.

A prototype is manufactured and given in Fig. 7(a), a base station antenna can be realised by using multiple subarrays shown in Fig. 7(b). In Fig. 7(a), a right-hand circular polarised (RHCP) array is mirrored to form a left-hand circular polarised array (LHCP), one subarray can be used to transmit the RHCP waves and another array can be used to receive the LHCP waves, or vice versa. This structure could have a very good spatial diversity with better isolation which will be evaluated experimentally in the future work.

ACKNOWLEDGEMENT

This work was supported in part by the National Natural Science Foundation of China (61701224 and 61601219) and Nature Science Foundation of Jiangsu Province (BK20160804). The work of T. H. Loh was supported in part by the 2017 – 2020 National Measurement System Programme of the UK government's Department for Business, Energy and Industrial Strategy (BEIS), under Science Theme Reference EMT17 of that Programme and in part by the EU project MET5G – Metrology for 5G communications (this project has received funding from the EMPIR programme co-financed by the Participating States and from the European Union's Horizon 2020 research and innovation programme), under EURAMET Reference 14IND10.

REFERENCES

[1] K. –C. Huang and Z. Wang, "Millimeter-wave circular polarized beam-steering antenna array for gigabit wireless communications," IEEE Transactions on Antennas and Propagation, vol. 54, no. 2, pp. 743-746, Feb. 2006.
[2] Y. Gao, R. Ma, Y. Wang, Q. Zhang and C. Parini, "Stacked patch antenna with dual-polarization and low mutual coupling for massive MIMO," IEEE Transactions on Antennas and Propagation, vol. 64, no. 10, pp. 4544-4549, Oct. 2016.
[3] Z. Y. Zhang, G. Fu, P. Huang, L. Pan, S. L. Zuo and J. Chen, "A wideband circularly polarized antenna with pattern improvement," Proceedings of 2014 3rd Asia-Pacific Conference on Antennas and Propagation, Harbin, 2014, pp. 547-550.
[4] F. Yang, A. Yu, A. Elsherbeni, and J. Huang, "Single-layer multi-band circularly polarized reflectarray antenna: concept, design, and measurement," 2008 URSI General Assembly, Chicago, IL, Aug. 8-16, 2008.
[5] Z. Su, G. Ding, W. Lu, W. Chen, "Broadband circularly polarized CPW-fed slot antenna array for millimeter wave application," Microelectronics Journal, vol. 40, pp. 1192-1195, 2009.
[6] P. Bhartia, I. Bahl, R. Garg and A. Ittipiboon, Microstrip Antenna Design Handbook, Artech House, 2000.
[7] J. D. Kraus and R. J. Marhefka, Antennas, McGraw-Hill Education, 3rd ed, 2001.
[8] R. C. Hansen, Electrically Small, Superdirective and Superconducting Antennas, Wiley-Interscience, 2006.

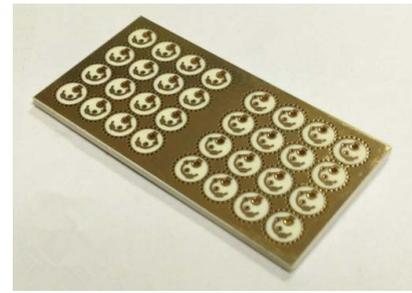
(a)

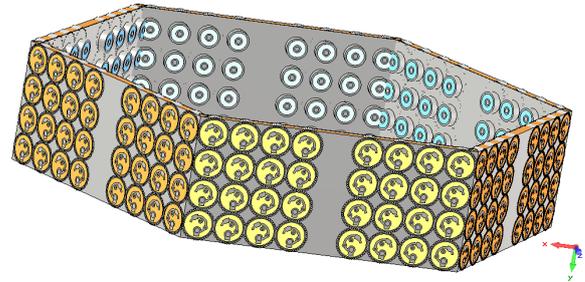
(b)

Fig. 7 (a) The manufactured subarray prototype, (b) a schematic plot of a MIMO base station.